\documentclass[prb,twocolumn,showpacs,preprintnumbers,amsmath,amssymb]{revtex4}

\usepackage{dcolumn}
\usepackage{bm}

\begin{document}

\title{Hedin's equations and enumeration of Feynman diagrams}

\author{L. G. Molinari} \email{luca.molinari@mi.infn.it}
\affiliation{Dipartimento di Fisica
dell'Universit\`a degli Studi di Milano\\ 
and I.N.F.N. sezione di Milano,
Via Celoria 16, I-20133
Milano}

\date{\today}

\begin{abstract}
Hedin's equations are solved perturbatively in zero dimension 
to count Feynman graphs for self-energy, polarization, propagator, effective 
potential and vertex function in a many-body theory of fermions with two-body 
interaction.  Counting numbers are also obtained in the GW approximation. 
\end{abstract}

\pacs{71.10.-w, 24.10.Cn, 11.10.Gh}

\maketitle

\section{Introduction}
Consider a system of fermions described by the Hamiltonian $H=\sum_i h(i)+
\sum_{i<j} v(i,j)$, where $h$ is a single-particle operator
and $v$ is the two-particle interaction. In general, to obtain dressed 
correlators that describe quasi-particles and collective excitations,  
one must sum over classes of diagrams of all orders of perturbation theory 
in $v$. The random phase approximation for the effective potential or the 
ladder expansion for the vertex and the T-matrix are two   
well known examples\cite{Mahan}. Iterative schemes are used in
conserving approximations for the self-energy\cite{Takada95, Schindlmayr01}.
The perturbative treatment of the electron gas requires a mixed approach, 
with partial 
resummations; a late step in the long history of the $r_s$-expansion of 
correlation energy is the evaluation of third order diagrams\cite{Endo}. 
In these cases, it might be useful, or at least interesting, to know in 
advance how many diagrams are required at each order of the approximation, 
and how many are left out. The counting problem is much simpler than the 
original one, and is
obtained by translating the recursive equations at hand to zero dimension. 
Properties other than loop content or perturbative numbers may also be 
evaluated: a notable example is diagram enumeration according to their 
topologies, that can be mapped to models of statistical mechanics on random 
lattices\cite{tHooft}.

In this report, I show how Feynman diagrams can be enumerated for the exact 
theory and its GW approximation. 
A useful starting point is the set of five Hedin's 
equations\cite{Hedin}
for the propagator $G(1,2)$, the effective potential $W(1,2)$, the proper 
self-energy $\Sigma (1,2)$, the polarization $\Pi(1,2)$, and the vertex 
$\Gamma(1;2,3)$:
\begin{eqnarray}
&& G(1,2)=g(1,2)+g(1,1')\Sigma(1',2')G(2',2)\label{G}\\
&& W(1,2)=v(1,2)+v(1,1')\Pi(1',2')W(2',2)\label{W}\\
&& \Sigma(1,2)=i\Gamma(2';1,1')G(1',2)W(2',2)\label{sigma}\\
&& \Pi(1,2)=-2i\Gamma(1;2',2'')G(2,2')G(2'',2)\label{polar}\\
&& \Gamma(1;2,3)=\delta(1,2)\delta(1,3)+\nonumber \\
&&\quad\Gamma(1;2',3')G(2'',2')G(3',3'')
\frac{\delta \Sigma (2,3)}{\delta G(2'',3'')}\label{gamma}
\end{eqnarray}
Repeated primed space-time and internal variables are hereafter understood 
to be summed or integrated. $g(1,2)$ is the Green function of the 
interacting system in the Hartree approximation, with exact particle 
density, so that Hartree-type insertion (tadpoles) are already accounted for.
In the vertex equation, the Bethe-Salpeter kernel\cite{BD} is 
specified as a functional derivative in the same quantities involved.  
Therefore, the five exact Hedin's equations are formally closed.
A different functional closure of Schwinger-Dyson equations, leading to 
diagrammatic recursion schemes, was developed by Kleinert et al.
\cite{Pelster02, Pelster03}.

It is useful to restate eq.(\ref{gamma}) with the Hartree propagator $g$ 
replacing the exact one, $G$:
\begin{eqnarray}
\Gamma (1;2,3)=\delta(1,2)\delta(1,3)+ 
g(1',1)g(1,1'')
\frac{\delta\Sigma(2,3)}{\delta g(1',1'')} \label{vertexg}
\end{eqnarray}
It states that vertex diagrams are obtained from self-energy 
ones through the insertion of a bare vertex in a Hartree propagator 
line, in all possible ways. A derivation is given in the end.
The formula can be advantageous when the self-energy is approximated by
some explicit expression of $g$ and $v$, which are the free 
functional variables of the many body problem. 

To cope with the vertex equation is a formidable task\cite{God,Oni}.
A strong non-trivial simplification is the GW approximation (GWA), 
where all corrections to the bare vertex $\Gamma^{(0)} (1;2,3)=
\delta(1,2)\delta(1,3)$ are ignored. Eqs. (\ref{G}-\ref{polar}), 
with $\Gamma_0$ replacing $\Gamma $, then become a closed set of integral 
equations \cite{Aul,Arya}. 

The upmost difficulty of the problem of solving Hedin's or GW equations, 
vanishes in zero dimension of space-time, where the perturbative solution 
in the interaction $v$ is a mean to count Feynman diagrams. The constraint 
of no Hartree-type insertions makes this approach much simpler than the usual 
one, based on the functional integral \cite{Cvi,Argy}. 
Counting numbers are also evaluated in the GWA, and the omission of vertex 
corrections makes the number of diagrams grow with a power law, instead 
of factorially.

\section{Counting Feynman diagrams}
In zero dimension of space-time, the four Hedin's equations 
(\ref{G}-\ref{polar}) become algebraic, with variables $g$ and $v$, 
and the functional 
derivative in the vertex equation (\ref{vertexg} ) is an ordinary one. After 
removing imaginary factors and replacing the factor $(-2)$, due to the 
fermionic loop and spin summation, with a parameter $\ell $ that counts 
the same loops, Hedin's equations are:
\begin{eqnarray}
  G=g+g\Sigma G,    \qquad  W=v+v\Pi W, \label{he12}
\end{eqnarray}
\begin{eqnarray}
 \Sigma=GW\Gamma,   \qquad  \Pi=\ell G^2\Gamma,\quad  
\Gamma=1+g^2 \frac{\partial \Sigma}{\partial g} \label{he345}
\end{eqnarray}
By searching solutions as series expansions in the variables $v$ and $\ell $, 
one obtains coefficients that count all Feynman graphs, with weight one, that 
contribute to a perturbative order (specified by the power of $v$), with a 
given number of fermionic loops (the power of $\ell $).
I begin by solving for the self-energy. Elimination of  $G$, $W$ and $\Pi$
gives 
\begin{eqnarray}
\Sigma(1-g\Sigma )^2 = gv\Gamma [1-(1-\ell)g\Sigma] \label{sigmagamma}
\end{eqnarray} 
It is useful to consider adimensional quantities, $\Sigma=gvs(x)$ and 
$x=g^2v$. After use of the equation for $\Gamma$ one obtains a differential
equation for $s(x)$ 
\begin{eqnarray}
s(1-xs)^2=(1+xs+2x^2 \frac{\partial s}{\partial x})[1-xs(1-\ell)]
\label{essex}
\end{eqnarray}
There is no standard solution of this (Abel) equation. The perturbative 
expansion is evaluated 
\begin{eqnarray}
&&\Sigma/vg = 1\,+\,(2+\ell)x\,+\,(10+9\ell+\ell^2)x^2 \,+\nonumber\\
&&\quad (74+91\ell+23\ell^2+\ell^3)x^3\,+\nonumber \\
&&\quad (706+1063\ell+416\ell^2+46\ell^3+
\ell^4)x^4\,+ \label{sigmaelle}\\
&&\quad (8162+14193\ell+7344\ell^2+1350\ell^3+80\ell^4+\ell^5)x^5+..
\nonumber  \\
&&=\,1+3x+20x^2+189x^3+2232 x^4+\nonumber \\
&&\quad 31130 x^5 +.. \label{sigmaelle1}
\end{eqnarray}
The last line corresponds to $\ell=1$. For example, in 
eq. (\ref{sigmaelle1})
we read that at order $v^3$ there are 20 self-energy diagrams; 
eq. (\ref{sigmaelle}) gives 
the more detailed information that 10 graphs come 
with no fermion loop, 9 with a single loop and 1 with two loops. 
At each perturbative order there is a single diagram with largest number of 
loops; the sum of such diagrams yields the ring approximation\cite{Mahan} 
$\Sigma_{r}=igW_r$, where the effective potential $W_r$ is evaluated
with ring insertions $\Pi_r=-2i g^2$.\par
The expansions for the vertex and the polarization are then 
obtained (I omit the propagator and the effective potential)
\begin{eqnarray}
&&\Gamma\,=\,1+x+(6+3\ell)x^2+(50+45\ell+5\ell^2)x^3+\nonumber \\
&&\quad (518+637\ell +161\ell^2 +7\ell^3)x^4 +\nonumber \\
&&\quad (6354+9567\ell+ 3744\ell^2+414\ell^3+9\ell^4)x^5 +..\\
&&=1+x+9x^2+100x^3+1323x^4+20088x^5+..
\end{eqnarray}
\begin{eqnarray}
&&\Pi/g^2\ell\,=\, 1+3x+ (15+5\ell)x^2+ (105+77\ell+7\ell^2)x^3
+\nonumber \\
&&\quad (945+1044\ell+234\ell^2+9\ell^3)x^4+\nonumber\\
&&\quad (10395+14784\ell+5390\ell^2+550\ell^3+11\ell^4)x^5+.. \\
&&=\, 1+3x+20x^2+189x^3+ 2232x^4+31130x^5 +..
\label{polelle1}
\end{eqnarray}
Note that the counting numbers for the $\ell=1$ expansions (\ref{sigmaelle1})
and (\ref{polelle1}) of the self-energy and the polarization are the same. 
This occurs also in Q.E.D. though with smaller counting numbers,
due to the cancellation of pairs of diagrams which differ by orientation of
a fermionic loop with an odd number of propagators (Furry's theorem) 
\cite{IZ}.
\par\vskip 0.5truecm

I now evaluate the counting numbers in the $GWA$, which corresponds to 
$\Gamma=1$ in eq.(\ref{sigmagamma}). The equation for the self-energy 
becomes algebraic cubic, with solution
\begin{eqnarray}
&&\Sigma_{GW}/vg\,=\, 1+(1+\ell)x+(2+4\ell+\ell^2)x^2+\nonumber \\
&&\quad (5+15\ell+9\ell^2+\ell^3)x^3+\nonumber \\
&&\quad (14+56\ell+56\ell^2+16\ell^3+\ell^4)x^4\,+ \\
&&\quad (42+210\ell+300\ell^2+150\ell^3+25\ell^4+\ell^5)x^5 +.. 
\nonumber\\
&&= \,1+2x+7x^2+30x^3+143 x^4+ 728 x^5 +.. 
\label{sgw}
\end{eqnarray}
For $\ell=1$, the cubic equation is $s(1-sx)^2=1$. By solving  
for the inverse function, $x(s)=s^{-1}-s^{-3/2}$, one locates a singular
point for $s'(x)$ in $x_c=4/27$. The finite radius of convergence
for the perturbative series (\ref{sgw}) in GWA, implies that
the number of self energy diagrams of order $n$ grows with the power law 
$(27/4)^n$. In GWA the vertex function is trivial, and the polarization 
$\Pi_{GW}=-2iG^2_{GW}$ is the GW-dressed ring diagram:  
\begin{eqnarray}
&&\Pi_{GW}/g^2\ell \, 
= \, 1\,+ 2x \,+ \, (5+2\ell)x^2\, +\, \nonumber \\
&&\quad (14+14\ell+2\ell^2)x^3\,+\nonumber\\
&&\quad (42+72\ell+27\ell^2+2\ell^3)x^4\,+\nonumber \\
&&\quad (132+330\ell+220\ell^2+44\ell^3+2\ell^4)x^5\,+\,..\\
&&=\, 1+2x+7x^2+30x^3+143x^4+728x^5+\,.. 
\end{eqnarray}
The $\ell =1$ expansions for the self-energy and the polarization 
are again the same, as one can show that $\Pi/g^2$ 
and $s$ solve the same cubic equation.

Conclusion: By translating to zero dimension a closed system of equations 
for many-body correlators, one may enumerate the Feynman diagrams involved.
I have shown this for the full theory of interacting fermions and
the GW approximation, with the rule that a line is a Hartree propagator.

\begin{acknowledgments}
This work was funded in part by the EU's 6th Framework Programme through 
 the NANOQUANTA Network of Excellence (NMP4-CT-2004-500198)
\end{acknowledgments}

\appendix
\section{}
I show that the vertex equation (\ref{gamma}) can be written in the form
(\ref{vertexg}); the simple proof is a modification of Hedin's construction, 
to which I refer. I recall the definition of the vertex function 
\begin{eqnarray}
\Gamma (1;2,3) = \delta(1,2)\delta(1,3) + 
\frac{\delta\Sigma(2,3)}{\delta V(1)}
\end{eqnarray}
The potential $V$ is the sum of an external potential (which is turned off at
the end) and the Hartree potential
 $\int d2 v(1,2)n(2)$, built with the exact particle density. Hedin's formula
is obtained by writing the functional derivative in $V$, through the chain 
rule, as a derivative in the exact Green function $G$. It is 
admissible to do so in terms of $g$:
\begin{eqnarray}
\frac {\delta\Sigma(2,3)}{\delta V(1)}=
\frac{\delta\Sigma(2,3)}{\delta g(1',1'')}
\frac{\delta g(1',1'')}{\delta V(1)} \nonumber
\end{eqnarray}
\noindent
From the Hartree equation $(i\partial(t'_1)-h(1')-V(1'))g(1',1'')=
\delta(1',1'')$, one evaluates the required functional derivative
$ \delta g(1',1'')/\delta V(1)= g(1',1)g(1,1'')$. This ends the proof.


\begin{thebibliography}{99}
\bibitem{Mahan}{G.D.Mahan, {\sl Many-Particle Physics}, (Kluwer 
Academic/Plenum Publishers, New York, 2000), 3rd ed.}
%
%
\bibitem{Takada95}{Y.Takada, Phys. Rev. B {\bf 52}, 12708 (1995).}
%
\bibitem{Schindlmayr01}{A.Schindlmayr, P.Garc\'{\i}a-Gonz\'alez and R.W.Godby,
Phys. Rev. B {\bf 64}, 235106-1 (2001).}
%
\bibitem{Endo}{T.Endo, M.Horiuchi, Y.Takada and H.Yasuhara, Phys. Rev. B
{\bf 59}, 7367 (1999).}
%
\bibitem{tHooft}{G. 'tHooft, Nucl. Phys. B {\bf 538}, 389 (1999).}
%
\bibitem{Hedin} {L. Hedin,  Phys. Rev. {\bf 139}, A796 (1965).} 
%
\bibitem{BD}{J.D.Bjorken and S.D.Drell, {\sl Relativistic Quantum Fields},
(McGraw-Hill, New York, 1965) ch.19.}
%
\bibitem{Pelster02}{A.Pelster, H.Kleinert and M.Bachmann, Ann. Phys. {\bf 297},
363 (2002).}
%
\bibitem{Pelster03}{A.Pelster and K.Glaum, Phys. Status Solidi B {\bf 237}, 72
(2003).}
%
\bibitem{God}{A.Schindlmayr and R.W.Godby,  Phys. Rev. Lett. {\bf 80}, 1702 
(1998).} 
%
\bibitem{Oni}{G.Onida, L.Reining and A.Rubio, Rev. Mod. Phys. {\bf 74}, 601 
(2002).}
%
\bibitem{Aul}{W.G.Aulbur, L.J{\"o}nsson and J.W.Wilkins, {\sl Solid State
Physics}, edited by H.Ehrenreich and F.Spaepen (Academic, New York, 2000),
vol. 54.} 
%
\bibitem{Arya}{F.Aryasetiawan and O.Gunnarsson, Rep. Progr. Phys. {\bf 61}, 237
(1998).}
%
\bibitem{Cvi}{P.Cvitanovic, B.Lautrup and R.B Pearson,  Phys. Rev. D {\bf 18},
1939 (1978).}
%
\bibitem{Argy}{E.N.Argyres, A.F.W. van Hameren, R.H.P. Kleiss, 
and C.G.Papadopoulos, Eur. Phys. J. C {\bf 19}, 567 (2001).}
%
\bibitem{IZ} {C.Itzykson and J.B.Zuber, {\sl Quantum Field Theory}, 
(McGraw Hill, New York, 1980) p.466.}
%
\end{thebibliography}
\end{document}